\documentclass[aps,showpacs,amssymb,amsfonts,twocolumn,superscriptadress]{revtex4}
\usepackage{amsmath,bbm,mathrsfs}
\usepackage{graphicx}
\usepackage{amsfonts}
\usepackage{amssymb}

\def\textbf#1{{\bf #1}}
\def\be{\begin{equation}}
\def\ee{\end{equation}}
\def\ben{\begin{eqnarray}}
\def\een{\end{eqnarray}}
\def\eea{\end{array}}
\def\bea{\begin{array}}
\newcommand{\ot}[0]{\otimes}
\newcommand{\Tr}[1]{\mathrm{Tr}#1}
\newcommand{\bei}{\begin{itemize}}
\newcommand{\eei}{\end{itemize}}
\newcommand{\ket}[1]{|#1\rangle}
\newcommand{\bra}[1]{\langle#1|}

\newcommand{\proj}[1]{\ket{#1}\!\bra{#1}}

\newcommand{\Ke}[1]{\big|#1\big\rangle}
\newcommand{\Br}[1]{\big< #1\big|}

\newcommand{\nmss}{\negmedspace\negmedspace}
\newcommand{\nmsss}{\negmedspace\negmedspace\negmedspace}

\begin{document}

\title{$W$--like bound entangled states and secure key distillation}

\author{Remigiusz Augusiak}
\email{remigiusz.augusiak@icfo.es} \affiliation{Faculty of Applied
Physics and Mathematics, Gda\'nsk University of Technology,
Narutowicza 11/12, 80--952 Gda\'nsk, Poland}
\affiliation{ICFO--Institute Ci\'encies Fot\'oniques,
Mediterranean Technology Park, 08860 Castelldefels (Barcelona),
Spain}

\author{Pawe{\l} Horodecki}
\email{pawel@mif.pg.gda.pl} \affiliation{Faculty of Applied
Physics and Mathematics, Gda\'nsk University of Technology,
Narutowicza 11/12, 80--952 Gda\'nsk, Poland}

\begin{abstract}
We construct multipartite entangled states with underlying
$W$--type structure satisfying positive partial transpose (PPT)
condition under any $(N-1)|1$ partition. Then we show how to
distill $N$--partite secure key form the states using two
different methods: direct application of local filtering and novel
random key distillation scheme in which we adopt the idea form
recent results on entanglement distillation. Open problems and
possible implications are also discussed.
\end{abstract}
\maketitle

\textbf{Introduction.} -- Quantum cryptography \cite{BB84,Ekert}
is an impressive information--theoretic application of quantum
physical laws in data security theory. The proofs
\cite{LoChau,ShorPreskill} of unconditional security of pioneering
quantum cryptographic protocol \cite{BB84} refer to the idea of
quantum privacy amplification \cite{QPA} based on entanglement
distillation protocol \cite{distillation}. This refers back to the
cryptographic protocol \cite{Ekert} which is based on shared pure
entanglement and is in fact the first explicit application of
entanglement in information theory. Since then we already know
that all correlation--based cryptographic protocols require
entanglement as a necessary resource \cite{Curty1}. While it was
natural to expect that distillation of pure entanglement is
necessary to cryptography, it happened that even nondistillable
entanglement known as a bound entanglement \cite{bound} may, at
least in some cases, be useful for cryptography \cite{KH0} with
the corresponding general entanglement--based cryptographic
paradigm going beyond entanglement distillation developed in
\cite{KH} (for recent interesting applications in security proofs
and physical analysis of security see Refs. \cite{RS,RB}).
Recently multipartite version of the latter has been worked out in
Refs. \cite{DoktoratRA,MultipartiteKey}. Especially in the latter
multipartite bound entanglement has been constructed based on
twisted GHZ--type of entanglement. Here we present a nonstandard
application  of the paradigm with a novel type of multipartite
bound entanglement, namely the one with underlying $W$-like
structure. We adopt here the idea of random distillation of
entanglement \cite{LoFortescue1,LoFortescue2} introducing the
notion of random distillation of secure key. The latter seems to
be much more efficient for the present states than the
concatenation of usual bipartite protocols with classical
postprocessing.

\textbf{$N$--partite noisy $W$--like states passing single--system
PPT test.} -- Below we provide a detailed construction of bound
entangled states, which exhibit the structure of noisy $W$ states,
where the latter are defined as $N$--qubit pure states of the form
\begin{equation}
\ket{W}=(1/\sqrt{N})(\ket{10\ldots0}+\ket{01\ldots
0}+\ldots+\ket{0\ldots 01}).
\end{equation}
We give a detailed proof that partial transposition with respect
to each single--party subsystem is positive.

Let us start by introducing the following matrices
\begin{equation}
Z_{D}=\sum_{i,j=0}^{D-1}u_{ij}\ket{ii}\!\bra{jj},\qquad
R_{D}=\sum_{i=0}^{D-1}\proj{ii},
\end{equation}
where $u_{ij}$ denotes elements of some unitary matrix $U_{D}$.
The sum of absolute values of all elements of $U_{D}$ will be
denoted by $\mathcal{U}_{D}$. For simplicity we can also assume
$U_{D}$ to be Hermitian. Now let us define
\begin{equation}\label{ConstrW1}
X_{D}^{(N)}=Z_{1,2}^{\Gamma_{2}}\ot \ldots \ot
Z_{i-1,i}^{\Gamma_{i}}\ot Z_{i,i+1}^{\Gamma_{i+1}}\ot \ldots \ot
Z_{N,1}^{\Gamma_{1}},
\end{equation}
where subscripts indicate that the matrix represents parts of
$i$--th and $j$--th party and $\Gamma_{j}$ stands for partial
transposition with respect to the subsystem belonging to $j$--th
party. Moreover, addition is modulo $N$. For instance
$Z_{1,2}^{\Gamma_{2}}$ is a part of quantum systems belonging to
the first and second party that must be transposed with respect to
the second party.

Let us now shortly discuss the properties of $X_{D}^{(N)}$.
Firstly, since $|Z_{i,i+1}|=|Z_{i,i+1}^{T}|=R_{D}$
$(i=1,\ldots,N)$ and
$\big|Z_{k-1,k}^{\Gamma_{k}}\big|=\sum_{i,j=0}^{D-1}|u_{ij}|\proj{ji}\;(\equiv
\mathcal{Z}_{k-1,k}),$
one concludes that
\begin{equation}\label{ConstrW4}
\left|\mathcal{X}_{D}^{(N)\Gamma_{i}}\right|=\bigotimes_{k=1}^{i-2}\mathcal{Z}_{k,k+1}\ot
R_{D}^{(2)}\ot R_{D}^{(2)}\ot
\bigotimes_{k=i+1}^{N}\mathcal{Z}_{k,k+1}.
\end{equation}
All the matrices $\big|\mathcal{X}_{D}^{(N)\Gamma_{i}}\big|$ are
diagonal and, as such, they are invariant under the action of
partial transposition. It is also clear that
$\big|X_{D}^{(N)}\big|=\bigotimes_{k=1}^{N}\mathcal{Z}_{k,k+1}$
which together with Eq. \eqref{ConstrW4} allows to infer that
\begin{equation}
\left\|X_{D}^{(N)\Gamma_{i}}\right\|_{1}=\mathcal{U}_{D}^{N-2}D^{2}
\quad \mathrm{and}\quad
\left\|X_{D}^{(N)}\right\|_{1}=\mathcal{U}_{D}^{N}.
\end{equation}
for any $i=1,\ldots,N$. Now we are prepared to present the
construction. For this purpose let us introduce
\begin{equation}
Y_{D}^{(N)}=\sum_{i=1}^{N}\left|X_{D}^{(N)\Gamma_{i}}\right|
\end{equation}
and denote by $\ket{\psi_{i}^{(N)}}$ ($\ket{\psi_{ij}^{(N)}}$)
pure $N$--qubit states in which $i$--th party ($i$--th and $j$--th
parties) posses $\ket{1}$ and the remaining parties have
$\ket{0}$. Let also $\mathcal{P}_{i}^{(N)}$ and
$\mathcal{P}_{ij}^{(N)}$ be projectors onto $\ket{\psi_{i}^{(N)}}$
and $\ket{\psi_{ij}^{(N)}}$, respectively.

Then we can consider the following class of states
\begin{eqnarray}\label{ConstrW8}
&&\hspace{-0.7cm}\varrho_{\mathsf{AA}'}^{(D,N)}=\frac{1}{\mathscr{N}_{D}^{(N)}}\left\{\sum_{\substack{i,j=1\\i\neq
j}}^{N}\Ke{\psi_{i}^{(N)}}\!\Br{\psi_{j}^{(N)}}\ot
X_{D}^{(N)}\right.\nonumber\\
&&\hspace{-0.5cm}+(N-1)\proj{0}^{\ot N}\ot Y_{D}^{(N)}+
\sum_{\substack{i,j=1\\i<j}}^{N}\mathcal{P}_{ij}^{(N)}\ot Y_{D}^{(N)}\nonumber\\
&&\hspace{-0.5cm}\left.+\sum_{\substack{i,j=1\\[1ex]\mbox{}}}^{N}
\mathcal{P}_{i}^{(N)}\ot
\left[(N-1)\left|X_{D}^{(N)}\right|+(N-2)Y_{D}^{(N)}\right]\right\},
\end{eqnarray}
where the normalization factor is given by
\begin{equation*}
\mathscr{N}_{D}^{(N)}=N\mathcal{U}_{D}^{N-2}\left[(N-1)\mathcal{U}_{D}^{2}+(D^{2}/2)(3N^{2}-3N-2)\right].
\end{equation*}
The subscripts $\mathsf{A}\equiv A_{1}\ldots A_{N}$ and
$\mathsf{A}'\equiv A_{1}'\ldots A_{N}'$ denote the {\it key part}
and {\it shield part} of the state. They are separated by the
tensor product visible in Eq. \eqref{ConstrW8}. "Everything" that
is on the left--hand side of this sign belongs to $\mathsf{A}$ and
everything on the right--hand side belongs to $\mathsf{A}'$.
Usually one considers the situation in which the $i$--th party has
two subsystems denoted here by $A_{i}$ and $A_{i}'$ (one from
$\mathsf{A}$ and one from $\mathsf{A}'$). However, in a more
general scenario we can also assume that the whole $\mathsf{A}'$
is held by some other but trusted party or even more trusted
parties.

Let us now check the positivity of partial transposition with
respect to the $i$--th subsystem. Straightforward algebra shows
that $\varrho_{\mathsf{AA}'}^{(D,N)\Gamma_{i}}$ is of the form
(Eq. \eqref{ConstrW10}).

\begin{widetext}

\begin{eqnarray}\label{ConstrW10}
\hspace{-0.0cm}\varrho_{\mathsf{AA}'}^{(D,N)\Gamma_{k}}&\nmsss\nmsss=\nmsss\nmsss&
\frac{1}{\mathscr{N}_{D}^{(N)}}\left\{
\left[\sum_{\substack{i=1\\i\neq k}}^{N}(\ket{0}^{\ot
N}\!\Br{\psi_{ik}^{(N)}} + \Ke{\psi_{ik}^{(N)}}\!\bra{0}^{\ot
N})\ot X_{D}^{(N)\Gamma_{k}}\right.\right.
\left.+(N-1)\proj{0}^{\ot N}\ot Y_{D}^{(N)}+\sum_{\substack{i=1\\i\neq k}}^{N}\mathcal{P}_{ik}^{(N)}\ot Y_{D}^{(N)}\right]\nonumber\\
&&\hspace{-1cm}+\left[(N-2)\sum_{\substack{i=1\\i\neq
k}}^{N}\mathcal{P}_{i}^{(N)}\ot Y_{D}^{(N)}
+\sum_{\substack{i,j=1\\i\neq j,\;i,j\neq
k}}^{N}\Ke{\psi_{i}^{(N)}}\!\Br{\psi_{j}^{(N)}}\ot
X_{D}^{(N)\Gamma_{k}}\right]
+\mathcal{P}_{k}^{(N)}\ot\left[(N-1)\left|X_{D}^{(N)}\right|+(N-2)Y_{D}^{(N)}\right]\nonumber\\
&&\left.+\sum_{\substack{i,j=1\\i<j,\; i,j\neq
k}}^{N}\hspace{-0.2cm}\mathcal{P}_{ij}^{(N)}\ot
Y_{D}^{(N)}\right\}.
\end{eqnarray}
\end{widetext}
To make the analysis simpler, some of the terms in the above were
grouped in square brackets. The positivity of the first and second
brackets follows straightforwardly from results of Ref.
\cite{MultipartiteKey} (see Lemma A.1). The remaining two terms
are positive as $Y_{D}^{(N)}\geq 0$.

Thus we showed that partial transposition with respect to any
single--party subsystem $A_{i}A_{i}'$ is positive. This indicates
that the states $\varrho_{\mathsf{AA}'}^{(D,N)}$ are bound
entangled provided that they are entangled. However, the latter
still needs to be shown. For this purpose below we discuss
cryptographical applicability of these states.

\textbf{Secure key distillation.} -- We prove that it is possible
to distill a nonzero amount of cryptographic key from the states
$\varrho_{\mathsf{AA}'}^{(D,N)}$. For this aim we show that one
can distill a bipartite secure key between any pair of parties of
$\varrho_{\mathsf{AA}'}^{(D,N)}$. Let us focus on the scenario in
which the remaining $N-2$ parties cooperate ''passively'', i.e.,
they perform no action but are trusted (do not cooperate with
Eve). In this case the distillable key\footnote{For definitions of
the bipartite and multipartite distillable key $C_{D}$ and $K_{D}$
the reader is referred to \cite{KH0,KH} and
\cite{DoktoratRA,MultipartiteKey}, respectively.} can only be
higher than in a scenario in which the remaining $N-2$ parties
would give some of their systems to Eve. Thus, for our purposes it
suffices to investigate the bipartite distillable key of the
states\footnote{The notation $\Tr_{\mathsf{A}\setminus\{k,l\}}$
means that we trace out the $\mathsf{A}$ subsystem except for
$A_{k}$ and $A_{l}$ subsystems. }
$\varrho_{A_{k}A_{l}\mathsf{A}'}^{(D,N)}=\Tr_{\mathsf{A}\setminus\{k,l\}}\varrho_{\mathsf{AA}'}^{(D,N)}$
for any $k\neq l$ (note that tracing out the subsystems may be
treated as giving them to Eve). As in what follows the additional
systems are not directly used in secure key distillation and the
remaining parties are trusted, we can considerably simplify the
analysis by applying the general bipartite cryptographical
paradigm studied in \cite{KH0,KH}. Indeed, we can even consider
the system $\mathsf{A}'$ as one distributed between $A_{k}$ and
$A_{l}$. However, it does not mean that the considered scenario is
only bipartite since the total protocol will consist of bipartite
protocols with different  pairs $\{ k, l \}$.

We investigate the bipartite distillable key using two methods.
The first one is quite simple application of local filtering,
while the second one, probably more efficient, bases on the ideas
of random distillation of entanglement given in Refs.
\cite{LoFortescue1,LoFortescue2}. Both protocols, are finally
concatenated with the Devetak--Winter (DW) protocol
\cite{DW1,DW2}.

We also simplify our considerations by imposing some constraints
on $U_{D}$, namely, we assume that all of its entries obey
$|u_{ij}|=1/\sqrt{D}$. An example of such a unitary Hermitian
matrix is the matrix $H^{\ot k}$, with $H$ being the Hadamard
matrix (here $D=2^{k}$). In this case $\mathcal{U}_{D}=D\sqrt{D}$
and what is important here
$\big\|X_{D}^{(N)}\big\|/\big\|Y_{D}^{(N)}\big\|=D/N$, which is
greater than one for sufficiently large $D$.

{\it Twistings and privacy squeezing.} In view of what was said
previously it suffices to restrict our considerations to the
distillation of bipartite secure key. The general cryptographical
paradigm of Refs. \cite{KH0,KH,KH1} is exactly what we need here.
Thus, below we recall some of its main ideas, namely, twistings
and the privacy squeezing \cite{KH} with its application in the
recent method \cite{KH1} of bounding the key form below. Possible
multipartite generalizations of the paradigm were studied in Refs.
\cite{DoktoratRA,MultipartiteKey}.

Let then $\varrho_{ABA'B'}$ denote some bipartite state with the
$AB$ ($A'B'$) part called the key (shield) part (notice once more
that in general in the considered scenario one does not have to
demand that the $A'B'$ part belong to the involved parties as it
may be in possession of some other trusted party). Now, let
$\mathcal{B}=\{\ket{ij}\}$ denote some product basis in the
Hilbert space corresponding to the $AB$ part. Then one defines the
{\it ccq state} $\varrho_{ABE}^{(\mathrm{ccq})}$ to be a state
that arises upon a measurement of the $AB$ part of a purification
$\ket{\psi_{ABA'B'E}}$ of $\varrho_{ABA'B'}$ in the product basis
$\mathcal{B}$ and tracing out the shield part $A'B'$ (in the usual
scenarios the shield part is treated then as a trivial subsystem).

Now, we define {\it twisting} (with respect to the basis
$\mathcal{B}$) to be the following operation:
\begin{equation}
U_{t}=\sum_{i,j}\proj{ij}\ot U_{ij}, \label{U-t}
\end{equation}
%
%
where in general $U_{ij}$ denote some isometries acting on the
$A'B'$ part. The important fact connected to twistings is that
$\varrho_{ABA'B'}$ and its twisted version
$U_{t}\varrho_{ABA'B'}U_{t}^{\dagger}$ have the same ccq state
(with respect to the same basis).

The last concept we would like to mention here is the so--called
{\it privacy squeezing}. Namely, by "rotating" the state
$\varrho_{ABA'B'}$ with some appropriately chosen twisting $U_t$
and then tracing out its shield part we get the {\it privacy
squeezed state}
$\widetilde{\varrho}_{AB}=\Tr_{A'B'}\big(U_{t}\varrho_{ABA'B'}U_{t}^{\dagger}\big)$.

Now, the method applied first in Ref. \cite{KH1} implies that
taking the purification $\ket{\widetilde{\Psi}_{ABE}}$ of the
latter and measuring it in the basis $AB$ produces the ccq state
with $C_{D}$ being a lower bound on distillable key of original
state $\varrho_{ABA'B'}$. Let us apply this technique carefully to
our example with general shield system $\mathsf{A}'$. Firstly, it
follows from Ref. \cite{KH} that in general
$K_{D}(\varrho_{A_{k}A_{l}\mathsf{A}'})=C_{D}(\ket{\psi_{A_{k}A_{l}\mathsf{A}'E}})\geq
C_{D}(\varrho_{A_{k}A_{l}E}^{(\mathrm{ccq})})$, where
$\ket{\psi_{A_{k}A_{l}\mathsf{A}'E}}$ stands for the purification
of $\varrho_{A_{k}A_{l}\mathsf{A}'}$, while
$\varrho_{A_{k}A_{l}E}^{(\mathrm{ccq})}$ denotes the ccq derived
according to the aforementioned prescription. On the other hand,
we can consider a twisted purification $\ket{\psi_{t}} \equiv
U_{t}\ot\mathbbm{1}_{E} \ket{\psi_{A_{k}A_{l}\mathsf{A}'E}}$. As
previously mentioned the ccq state (denoted as
$\sigma_{A_{k}A_{l}E}^{(\mathrm{ccq})}$) following this
purification is exactly the same as
$\varrho_{A_{k}A_{l}E}^{(\mathrm{ccq})}$ (in $\mathcal{B}$).
Finally, we can consider a worse situation from the point of
secure key distillation between the parties $A_{k}$ and $A_{l}$.
Namely giving now the $\mathsf{A}'$ subsystem we can only lower
the key. In other words, we can look at the twisted purification
$\ket{\psi_{t}}$ as coming from purifying only the $A_{k}A_{l}$
(with the whole system $E'=\mathsf{A}'E$ considered to be in Eve's
hands). In this way we have
$C_{D}(\sigma_{A_{k}A_{l}E}^{(\mathrm{ccq})})\geq
C_{D}(\widetilde{\varrho}_{A_{k}A_{l}E'}^{(\mathrm{ccq})})$, where
$\widetilde{\varrho}_{A_{k}A_{l}E'}^{(\mathrm{ccq})}$ denotes the
ccq state derived in this way. The last step is an application of
the Devetak--Winter protocol to
$\widetilde{\varrho}_{A_{k}A_{l}E}^{(\mathrm{ccq})}$ which gives
$C_{D}(\widetilde{\varrho}_{A_{k}A_{l}E}^{(\mathrm{ccq})})\geq
I(A_{k}\!:\!A_{l})-I(A_{k}\!:\!E),$
where the quantities\footnote{The quantum mutual information
$I(A\!:\!B)$ is defined for $\varrho_{AB}$ as
$I(A\!:\!B)=S(\varrho_{A})+S(\varrho_{B})-S(\varrho_{AB})$ with
$S$ denoting the von Neumann entropy.} $I(A_{k}\!:\!A_{l})$ and
$I(A_{k}\!:\!E)$ are calculated for respective bipartite
reductions of
$\widetilde{\varrho}_{A_{k}A_{l}E}^{(\mathrm{ccq})}$. The
conclusion following this analysis is that
$K_{D}(\varrho_{A_{k}A_{l}\mathsf{A}'})\geq
C_{D}(\widetilde{\varrho}_{A_{k}A_{l}E}^{(\mathrm{ccq})})$ and
therefore in what follows we can restrict to the analysis of
distillable key of
$\widetilde{\varrho}_{A_{k}A_{l}E}^{(\mathrm{ccq})}$. In other
words we need to take privacy--squeezed
$\widetilde{\varrho}_{A_{k}A_{l}}^{(D,N)}$ version of
$\varrho_{A_{k}A_{l}\mathsf{A}'}^{(D,N)}$ and analyze lower bounds
on the distillable key of its ccq state.

Note also that any filtering operation diagonal in $\mathcal{B}$
and performed on the key part of the state commutes with the
privacy squeezing operation with respect to the same basis. This
allows to perform local filters on the privacy--squeezed state
instead of the initial one $\varrho_{\mathsf{AA}'}^{(D,N)}$.

{\it Direct application of local filters.} Without loss of
generality we can assume $\mathcal{B}$ to be the standard basis in
$\mathbb{C}^{2}\ot\mathbb{C}^{2}$. Then we can derive the
privacy--squeezed state of an arbitrary state
$\varrho_{A_{k}A_{l}\mathsf{A}'}$ ($k\neq l$). Choosing in
(\ref{U-t}) $U_{01}^{\dagger}$ and $U_{10}^{\dagger}$ to be
unitary matrices from the singular--value decomposition of
$X_{D}^{(N)}$ and $U_{00}=U_{11}=\mathbbm{1}_{D^{2N}}$, we get
after some calculations from Eq. \eqref{ConstrW8} that

\begin{eqnarray}\label{ConstrW15}
\widetilde{\varrho}_{A_{k}A_{l}}^{(D,N)}=
\frac{1}{\widetilde{\mathscr{N}}_{D}^{(N)}}\left[
\begin{array}{cccc}
\alpha_{D,N} & 0 & 0 & 0 \\
0 & \beta_{D,N} & D & 0 \\
0 & D & \beta_{D,N} & 0 \\
0 & 0 & 0 & N
\end{array}
\right],
\end{eqnarray}
where
$\widetilde{\mathscr{N}}_{D}^{(N)}=N[(N-1)D+(3N^{2}-3N-2)/2]$,
$\alpha_{D,N}=(N-2)(N-1)D+[(3N^{2}-11N+12)/2]N$, and
$\beta_{D,N}=(N-1)D+2(N-2)N$.

\begin{figure}[h!]
\centering(a)\includegraphics[width=6cm]{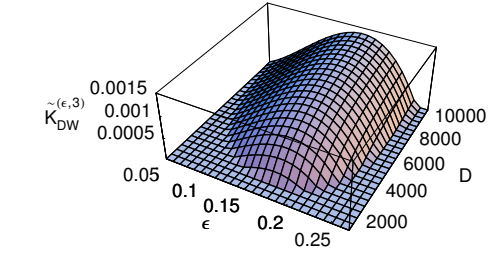}\\
\centering(b)\includegraphics[width=6cm]{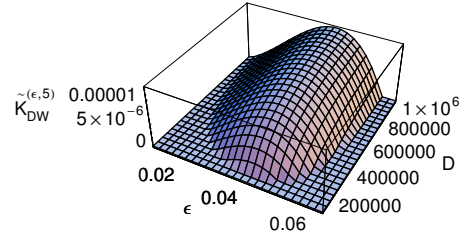} \caption{The
dependence of $\widetilde{K}_{DW}^{(\epsilon,N)}$ on the
parameters $\epsilon$ and $D$ for two different values of $N$,
namely, $N=3$ (a) and $N=5$ (b). Zero is put whenever the plotted
function is less than zero. Also, for clarity, the function is
plotted as if it were continuous in $D$. It is clear from both the
plots that the number of parties lowers the plotted
function.}\label{ConstrWFig3}
\end{figure}

Since $\alpha_{D,N}$ considerably dominates the remaining entries
of $\widetilde{\varrho}_{A_{k}A_{l}}^{(D,N)}$, the DW protocol
does not apply here. However, using some local
filters\footnote{Physically the filters are performed on
subsystems of the key part of $\varrho_{\mathsf{AA}'}^{(D,N)}$ by
respective parties, however as they commute with the privacy
squeezing we can mathematically perform it on
$\widetilde{\varrho}_{A_{k}A_{l}}^{(D,N)}$.} we can change the
respective entries. So, let us consider the filter
$V_{\epsilon}=\mathrm{diag}[\epsilon,1]$ $(0\leq \epsilon\leq 1)$
and let the $k$--th and $l$--th party apply it. This with
probability
$q_{D,N}^{(\epsilon)}=\Tr\big(V^{\dagger}_{\epsilon}V_{\epsilon}\ot
V^{\dagger}_{\epsilon}V_{\epsilon}\,\widetilde{\varrho}_{A_{k}A_{l}}^{(D,N)}\big)$
brings $\widetilde{\varrho}_{A_{k}A_{l}}^{(D,N)}$ to the following
state
\begin{eqnarray}\label{StateFiltering}
\hspace{-0.5cm}\overline{\varrho}_{A_{k}A_{l}}^{(D,N,\epsilon)}
&\nmss=\nmss&\frac{\epsilon^{2}}{\overline{\mathscr{N}}_{D,N}^{(\epsilon)}}
\left[
\begin{array}{cccc}
\alpha_{D,N}\epsilon^{2} & 0 & 0 & 0 \\
0 & \beta_{D,N} & D & 0 \\
0 & D & \beta_{D,N} & 0 \\
0 & 0 & 0 & \frac{N}{\epsilon^{2}}
\end{array}
\right]
\end{eqnarray}
with
$\overline{\mathscr{N}}_{D,N}^{(\epsilon)}=\alpha_{D,N}\epsilon^{4}+
2\beta_{D,N}\epsilon^{2}+N$. According to the previous
prescription what we need now is to bound from below the
distillable key of the ccq state (in $\mathcal{B}$) of
$\widetilde{\varrho}_{A_{k}A_{l}}^{(D,N)}$. For this purpose we
can firstly find a lower bound on $C_{D}$ of
$\overline{\varrho}_{A_{k}A_{l}E}^{(\mathrm{ccq},\epsilon)}$ using
the DW protocol, where by
$\overline{\varrho}_{A_{k}A_{l}E}^{(\mathrm{ccq},\epsilon)}$ we
denoted the ccq state\footnote{For the sake of clearity, we do not
provide explicit form of the purification of
$\overline{\varrho}_{A_{k}A_{l}}^{(D,N,\epsilon)}$ and ccq state
$\overline{\varrho}_{A_{k}A_{l}E}^{(\mathrm{ccq},\epsilon)}$.}
corresponding to the  output of the filtering, i.e.,
$\overline{\varrho}_{A_{k}A_{l}}^{(D,N,\epsilon)}$. Secondly, as
local filtering is a stochastic operation, multiplying the latter
with the success probability $q_{D,N}^{(\epsilon)}$ we get the
desired result. This, however, according to the discussion above
allows us to write
\begin{eqnarray}\label{kolejnyBound}
K_{D}(\varrho_{A_{k}A_{l}\mathsf{A}'}^{(D,N)})\geq
q_{D,N}^{(\epsilon)}\left[I(A_{k}\!:\!A_{l})
-I(A_{k}\!:\!E)\right],
\end{eqnarray}
where both $I$ are calculated from reductions of
$\overline{\varrho}_{A_{k}A_{l}E}^{(\mathrm{ccq},\epsilon)}$. The
behaviour of the right--hand side (denoted by
$\widetilde{K}_{DW}^{(\epsilon,N)}$) of Eq. \eqref{kolejnyBound}
as a function of the filter parameter $\epsilon$ and the dimension
$D$ is plotted in Fig. \ref{ConstrWFig3} for $N=3$ and $N=5$.
Despite the rather small values of
$\widetilde{K}_{DW}^{(\epsilon,N)}$ and the large dimension $D$ of
$A_{i}'$, it is clear from Fig. \ref{ConstrWFig3} that one may
distill a nonzero amount of bipartite key from the states
$\varrho_{\mathsf{AA}'}^{(D,3)}$ and
$\varrho_{\mathsf{AA}'}^{(D,5)}$.
%

Finally, we need to show that indeed the possibility of secure key
distillation between any pair of parties of
$\varrho_{\mathsf{AA}´}^{(D,N)}$ leads to the distillation of a
genuine multipartite secure key among all the parties. For this
purpose notice firstly that in the general case of an $N$--partite
state it suffices to have a bipartite secure key among pairs
$A_{i}A_{i+1}$ $(i=1,\ldots,N-1)$. Secondly, let us assume that
each such pair distills a secure key at a rate $r$. Then one
concludes that in such a configuration all the parties can distill
multipartite key at a rate at least $r/(N-1)$. Since we showed
that in the case of our states $r$ is nonzero, the multipartite
distillable key of $\varrho_{\mathsf{AA}'}^{(D,N)}$ is nonzero, at
least in the cases of $N=3,5$.

{\it Alternative approach: the idea of random distillation of
secure key.} -- Now, basing on the very recent results of Lo and
Fortescue \cite{LoFortescue1,LoFortescue2}, we consider a little
bit more sophisticated way of the bipartite key distillation from
$\varrho_{\mathsf{AA}'}^{(D,N)}$. For simplicity we focus here
only on the case of $N=3$, however, generalization to more parties
is straightforward.

Let us then consider the following POVM
$\mathscr{V}_{\epsilon}=\mathrm{diag}[\sqrt{1-\epsilon^{2}},1]$
and $\mathscr{W}_{\epsilon}=\mathrm{diag}[\epsilon,0]$ $(0\leq
\epsilon\leq 1)$.
It is clear that
$\mathscr{V}_{\epsilon}^{\dagger}\mathscr{V}_{\epsilon}
+\mathscr{W}_{\epsilon}^{\dagger}\mathscr{W}_{\epsilon}=\mathbbm{1}_{2}$,
where $\mathbbm{1}_{2}$ denotes the $2\times 2$ identity. Each of
the parties applies this POVM to their "nonprimed" subsystems
$A_{i}$ $(i=1,2,3)$. Now, we divide the possible outcomes into
three groups. The first one contains a single element, namely, a
result of application of $\mathscr{V}_{\epsilon}^{\ot 3}$. The
second group contains the outcome of the application of
$\mathscr{V}_{\epsilon}^{\ot 2}\ot \mathscr{W}_{\epsilon}$ and two
other outcomes being permutations of $\mathscr{V}_{\epsilon}$ and
$\mathscr{W}_{\epsilon}$ in $\mathscr{V}_{\epsilon}^{\ot 2}\ot
\mathscr{W}_{\epsilon}$. Finally, the third group consists of the
remaining outcomes. The results from the second group are treated
as a success since they lead to secure key distillation. On the
contrary any result from the third group is considered as a
failure as the resulting state has a separable structure with
respect to the key part. In the case when the obtained result
belongs to the second or third group, the protocol stops. On the
other hand, when the result belongs to the first group we have to
repeat our protocol as the obtained result keeps the structure of
the initial state.

Let us now pass to the protocol. Assume that the parties repeat
the measurement $M$ times, but in such way that in each round the
value of $\epsilon$ in the definition of POVM differs. Precisely,
following Ref. \cite{LoFortescue1} we utilize
$\epsilon_{i}=1/\sqrt{1+i}$, however, in a reversed order, i.e.,
in the first round we take $\epsilon_{M}=1/\sqrt{1+M}$ and in the
last one $\epsilon_{1}=1/2$.

Taking into account a single success outcome
$\mathscr{V}_{\epsilon}^{\ot 2}\ot \mathscr{W}_{\epsilon}$
(corresponding to the secure key distillation between the first
and second party), the state after $M$ measurements is of the form
$\rho_{\mathsf{AA}'}^{(D,M)}=G_{D}^{(M)}/\Tr(G_{D}^{(M)})$,
where\footnote{By $\widetilde{\mathscr{V}}_{\epsilon}$ and
$\widetilde{\mathscr{W}}_{\epsilon}$ we denoted POVM operators
extended by the identity on $\mathsf{A}'$ subsystems.}
\begin{eqnarray}\label{G}
&&G_{D}^{(M)}=\widetilde{\mathscr{V}}_{\epsilon_{M}}^{\ot 2}\ot
\widetilde{\mathscr{W}}_{\epsilon_{M}}\varrho_{\mathsf{AA}'}^{(D,3)}
\widetilde{\mathscr{V}}_{\epsilon_{M}}^{\ot 2}\ot\widetilde{\mathscr{W}}_{\epsilon_{M}}\nonumber\\
&&+\widetilde{\mathscr{V}}_{\epsilon_{M-1}}\widetilde{\mathscr{V}}_{\epsilon_{M}}\ot
\widetilde{\mathscr{V}}_{\epsilon_{M-1}}\widetilde{\mathscr{V}}_{\epsilon_{M}}\ot
\widetilde{\mathscr{W}}_{\epsilon_{M-1}}\widetilde{\mathscr{V}}_{\epsilon_{M}}\nonumber\\
&&\;\;\;\times
\varrho_{\mathsf{AA}'}^{(D,3)}\widetilde{\mathscr{V}}_{\epsilon_{M}}\widetilde{\mathscr{V}}_{\epsilon_{M-1}}\ot
\widetilde{\mathscr{V}}_{\epsilon_{M}}\widetilde{\mathscr{V}}_{\epsilon_{M-1}}\ot
\widetilde{\mathscr{V}}_{\epsilon_{M}}\widetilde{\mathscr{W}}_{\epsilon_{M-1}}\nonumber\\
&&\vdots\nonumber\\
&&+\widetilde{\mathscr{V}}_{\epsilon_{1}}\ldots
\widetilde{\mathscr{V}}_{\epsilon_{M}}\ot
\widetilde{\mathscr{V}}_{\epsilon_{1}}\ldots
\widetilde{\mathscr{V}}_{\epsilon_{M}}\ot
\widetilde{\mathscr{W}}_{\epsilon_{1}}\widetilde{\mathscr{V}}_{\epsilon_{2}}\ldots
\widetilde{\mathscr{V}}_{\epsilon_{M}}
\varrho_{\mathsf{AA}'}^{(D,3)}\nonumber\\
&&\;\;\;\times\widetilde{\mathscr{V}}_{\epsilon_{M}}\ldots
\widetilde{\mathscr{V}}_{\epsilon_{1}}\ot
\widetilde{\mathscr{V}}_{\epsilon_{M}}\ldots
\widetilde{\mathscr{V}}_{\epsilon_{1}}\ot
\widetilde{\mathscr{V}}_{\epsilon_{M}}\widetilde{\mathscr{V}}_{\epsilon_{2}}\ldots
\widetilde{\mathscr{W}}_{\epsilon_{1}}.
\end{eqnarray}

The probability of appearance of $\rho_{\mathsf{AA'}}^{(D,M)}$ is
given by $q_{D}^{(M)} =(2M^{2}(D+4)+M(2D+7))/6(D+4)(M+1)^{2}.$

Let us briefly explain Eq. \eqref{G}. The first term corresponds
to the success obtained in the first round of the protocol, while
the second term is responsible for the outcome from the first
group obtained in the first round and the success obtained in the
second round. The remaining terms may be derived in an analogous
way.

Since we chose the success outcome corresponding to the key
distillation between parties $A_{1}$ and $A_{2}$ we can trace the
key part of the last party of $\rho_{\mathsf{AA'}}^{(D,M)}$,
getting the state $\rho_{A_{1}A_{2}\mathsf{A'}}^{(D,M)}$. As we
are interested in application of the DW protocol we can apply the
privacy squeezing with the same twisting operation $U_{t}$ as in
previous subsection, which effectively removes the shield part and
produces finally
\begin{equation}\label{StanyWrownanie}
\widetilde{\rho}_{A_{1}A_{2}}^{(D,M)}=\frac{1}{\mathcal{G}_{D}^{(M)}}
\left[
\begin{array}{cccc}
2\frac{2M+1}{M+1} & 0 & 0 & 0 \\
0 & 2D+3 & D & 0 \\
0 & D & 2D+3 & 0 \\
0 & 0 & 0 & 6
\end{array}
\right],
\end{equation}
where $\mathcal{G}_{D}^{(M)}=2[2M(D+4)+2D+7]/(M+1)$.

The remaining two success outcomes of the POVM (corresponding to
$\mathscr{V}_{\epsilon}\ot
\mathscr{W}_{\epsilon}\ot\mathscr{V}_{\epsilon}$ and
$\mathscr{W}_{\epsilon}\ot\mathscr{V}_{\epsilon}^{\ot 2}$) lead
after $M$ rounds to exactly the same two--qubit states as in Eq.
\eqref{StanyWrownanie}, however, shared by the parties $A_{1}$ and
$A_{3}$, and $A_{2}$ and $A_{3}$, respectively. Also,
probabilities of obtaining the respective states are the same and
equal to $q_{D}^{(M)}$. Let us notice also that in the asymptotic
limit $M\to \infty$ the probability $q_{D}^{(M)}$ tends to
one--third. It means that taking into account all the three
success outputs we are sure that in the limit of $M\to\infty$ the
secure bit will be shared by one of the pairs of parties.

\begin{figure}[!ht]
\centering\includegraphics[width=5cm]{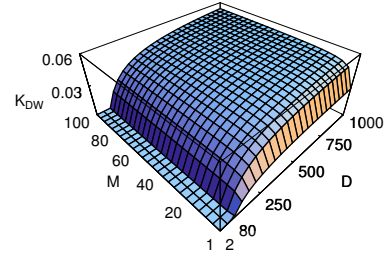} \caption{The
dependence of $K_{DW}$ on $M$ and $D$. Zero is put whenever the
plotted function is less than zero and, for convenience, it is
presented as a function of continuous $M$ and $D$. Interestingly,
the nonzero values appear at about $D=80$, while in the analogous
plot (Fig. \ref{ConstrWFig3}a) nonzero values are from about
$D=2000$. Moreover, it is clear that in the case of the random
protocol the distillable key is bounded by larger values.
Consequently, it is very reasonable to suspect that by using the
random protocol one can distill more secure key.} \label{Random2}
\end{figure}

Finally, in the same way as previously we get
\begin{eqnarray}\label{lowerbound}
K_{D}(\varrho_{A_{1}A_{2}\mathsf{A'}}^{(D,3)})\geq
q_{D}^{(M)}\left[I(A_{1}\!:\!A_{2}) -I(A_{1}\!:\!E)\right],
\end{eqnarray}
where $I$ are calculated for reductions of the ccq state
$\widetilde{\rho}_{A_{1}A_{2}E}^{(\mathrm{ccq})}$ (in
$\mathcal{B}$) of $\widetilde{\rho}_{A_{1}A_{2}}^{(D,M)}$. The
behavior of the function appearing on the right--hand side of the
above, i.e., the difference between mutual information multiplied
by $q_{D}^{(M)}$ (denoted by $K_{DW}$) is presented in Fig.
\ref{Random2}. On the other hand, one may prove analytically that
it is possible to get a secure key from
$\widetilde{\rho}_{A_{1}A_{2}E}^{(\mathrm{ccq})}$. Namely, notice
that the limit of $\widetilde{\rho}_{A_{1}A_{2}}^{(D,M)}$ with
$D\to\infty$ has nonzero $K_{D}$ by the DW protocol and thus, by
continuity of the involved functions there must exist such $D$
that $\widetilde{\rho}_{A_{1}A_{2}E}^{(\mathrm{ccq})}$ has also
nonzero $K_{D}$.

To finish our considerations we discuss what rates of multipartite
key are achievable within the described method. We already know
that to get the multipartite key it suffices to have secure key
between some properly chosen parties. However in previous cases we
needed to divide protocol into separate $N-1$ bipartite
deterministic protocols,  while here we get different bipartite
keys in {\it one} deterministic protocol. This suggests that we
can think a little bit clever while estimating the multipartite
key rate here.

For this purpose let us focus on the three--partite case and and
consider the rates $r_{1},r_{2},r_{3}$, where $r_{1}$ is a key
rate between $A_{1}$ and $A_{2}$ and so on. Assume that all the
rates $r_i$ are positive and form the triangle inequality. In this
case there exist a triple of positive numbers $a,b,c$ such that
$(r_{1},r_{2},r_{3})=(a+b,b+c,c+a)$, it is not difficult to
conclude that the rate of multipartite key may be lower bounded by
$a+b+c=(r_{1}+r_{2}+r_{3})/2$. For
$\varrho_{\mathsf{AA}'}^{(D,N)}$ all rates $r_{i}$ are equal to
$K_{DW}$ and the obtainable rate of multipartite key is
$(3/2)K_{DW}$.

{\it Distillation of a "truly" random secure key -- is this
possible for bound entangled states?} In the previous section we
have considered random distillation of a secure key. In this
process all the parties have to cooperate. Indeed the performances
of the protocol have been estimated with privacy squeezing
involving {\it all} "primed" parties as a shield part. This means
that during the protocol the passive parties (like $A_{3}$ in the
analysis from the previous subsection) were trusted in the sense
that keep their "primed" subsystems ($A_{3}'$) and do not give it
to Eve.

This means that effectively the bipartite key between $A_{1}$ and
$A_{2}$ can only at this early stage get correlated to the third
party. But what if we were interested in a {\it true random
bipartite key}, i.e., such that after the random bipartite
protocol the (random) bipartite key is secure not only against Eve
but also against other parties?

This may be also related to a two--stages protocol: $N$ trusted
parties representing some public company are given $N$--partite
state and distill such random key in a way that it is truly
bipartite. After that, $N$ different parties come and use that key
having all the bipartite secure communications guaranteed.

Note that this kind of random secure key would share with
entanglement the monogamy property. The natural question is which
multipartite bound entangled states can lead to the key with such
a property. Preliminary analysis of our $W$--like states seems to
suggests that it is impossible to get such key form our states.

\textbf{Discussion.} -- We have provided a construction of novel
multipartite bound entangled states with underlying $W$--type
structure. The states satisfy PPT test for any $(N-1)|1$
partition. We have analyzed distillation of a secure key form the
states in two different ways. The first one is based on the usual
bipartite filtering--based protocol followed by the DW scheme. The
second one involves random distillation of secure key. Though we
have not proven the optimality of the protocols, the present
results suggest that, as in the entanglement distillation, the
random distillation of a secure key may be much more efficient in
the distillation of a multipartite cryptographic key in cases when
one deals with underlying $W$--type structure. However, since the
present states are the first bound entangled states of this type,
still further analysis is necessary. One also expects that bound
entanglement of other multipartite types like graph states
\cite{graph} may be also constructed and found to be useful in
quantum cryptography. It is interesting to address this type of
questions in the context of recently discovered thermal bound
entanglement in quantum arrays and lattices \cite{thermal}. On the
other hand, one may ask about the distillation of a quantum key in
a modified sense: this would be the "truly" random bipartite key
in the sense that bipartite cryptographic correlations were secure
not only against Eve but also against all the remaining parties.
Note that, of course this is possible in case of some free
entangled states: Lo--Fortesque protocol followed by classical
measurement of entangled pairs provides naturally such key. Here
the natural question arises which bound entangled states lead to
such key.

Another natural question concerns the relation of the present
results to quantum channels capacities. Indeed the present states
as well as the states from \cite{MultipartiteKey} may be
immediately used to generate a quantum channel (with $k$ senders
and $n-k$ receivers). It is interesting that while the one--sender
channels created from the GHZ--type states \cite{MultipartiteKey}
have strictly positive one--way multipartite privacy capacity
${\cal P}$ nonzero (due to generalized DW protocol) it seems to be
rather unlikely that the channels based on the present $W$-like
states have that property. Still, in context of fascinating and
still uncovered role of privacy in the recently discovered
superactivation effect of quantum bipartite capacity
\cite{Science}, and especially in the context of multipartite
superactivation and activation of quantum capacities and
entanglement (see \cite{ActivationSuperactivationPapers}), further
analysis of quantum channels based on the present states seems to
be interesting.

\acknowledgments The work is supported by  EU Integrated Project
SCALA and by LFPPI network. R. A. gratefully acknowledges the
support from Ingenio 2010 QOIT and Foundation for Polish Science.


\begin{thebibliography}{0}

\bibitem{BB84}
Bennett C. H. and Brassard G., {\it Quantum cryptography: Public
key distribution and coin tossing, in Proceedings of the IEEE
International Conference on Computers, Systems and Signal
Processing, Bangalore, India, December, 1984} (IEEE Computer
Society Press, New York) p. 175.

\bibitem{Ekert}A. K. Ekert, Phys. Rev. Lett. {\bf 67}, 661 (1991).

\bibitem{ShorPreskill}P. W. Shor and J. Preskill,
Phys. Rev. Lett. {\bf 85}, 441 (2000).

\bibitem{LoChau}H.--K. Lo and H. F. Chau, Science {\bf 283}, 2050 (1999).


\bibitem{QPA}D. Deutsch, A. K. Ekert, R. Jozsa, C. Macchiavello, S. Popescu, and A.
Sanpera, Phys. Rev. Lett. {\bf 77}, 2818 (1996).

\bibitem{distillation}C. H. Bennett, G. Brassard, S. Popescu, B. Schumacher, J. A. Smolin and W. K. Wootters,
Phys. Rev. Lett. {\bf 76}, 722 (1996).



\bibitem{Curty1}M. Curty, M. Lewenstein and N. L\"utkenhaus, Phys. Rev. Lett. {\bf 92}, 217903 (2004).














\bibitem{bound}M. Horodecki, P. Horodecki, and R. Horodecki,
Phys. Rev. Lett. {\bf 80}, 5239 (1998).

\bibitem{KH0}K. Horodecki, M. Horodecki, P. Horodecki and J.
Oppenheim, Phys. Rev. Lett. {\bf 94}, 160502 (2005).

\bibitem{KH}K. Horodecki, M. Horodecki, P. Horodecki and J.
Oppenheim, IEEE Trans. Inf. Theor. {\bf 55}, 1898 (2009).



\bibitem{RS}J. M. Renes, G. Smith, Phys. Rev. Lett. {\bf 98}, 020502 (2007).


\bibitem{RB}J. M. Renes, J.--Ch. Boileau, Phys. Rev. A {\bf 78}, 032335 (2008).

\bibitem{DoktoratRA}R. Augusiak, {\it On the distillation of secure key form
multipartite entangled quantum states}, PhD thesis, Gda\'nsk,
2008.

\bibitem{MultipartiteKey}R. Augusiak and P. Horodecki,
{\it Multipartite secret key distillation and bound entanglement
}, arXiv:0811.3603, in press in Phys. Rev. A.





\bibitem{LoFortescue1}H.--K. Lo and B. Fortescue,
Phys. Rev. Lett. {\bf 98}, 260501 (2007).


\bibitem{LoFortescue2}H.--K. Lo and B. Fortescue,
Phys. Rev. A {\bf 78}, 012348 (2008).








\bibitem{DW1}I. Devetak and A. Winter, Phys. Rev. Lett. {\bf 93}, 080501 (2004).

\bibitem{DW2}I. Devetak and A. Winter, Proc. R. Soc. Lond. A {\bf 461}, 207 (2005).

\bibitem{KH1}K. Horodecki, \L{}. Pankowski, M. Horodecki and P.
Horodecki, IEEE Trans. Inf. Theory {\bf 54}, 2621 (2008).





\bibitem{graph}R. Raussendorf, D. Browne and
H.--J. Briegel, Phys. Rev. A {\bf 68}, 022312 (2003).

\bibitem{thermal}G. T\'oth, C. Knapp, O. G\"uhne and H. J. Briegel,
Phys. Rev. Lett. {\bf 99}, 250405 (2007); A. Ferraro, D.
Cavalcanti, A. Garc\'ia--Saez and A. Ac\'in, Phys. Rev. Lett. {\bf
100}, 080502 (2008).


\bibitem{Science}G. Smith and J. Yard, Science {\bf 321}, 1812 (2008).


\bibitem{ActivationSuperactivationPapers}P. W. Shor, J. A. Smolin and A. V. Thapliyal,
Phys. Rev. Lett. {\bf 90}, 107901 (2003); W. D\"ur, J. I. Cirac
and P. Horodecki, Phys. Rev. Lett {\bf 93}, 020503 (2004); \L{}.
Czekaj and P. Horodecki, Phys. Rev. Lett. {\bf 102}, 110505
(2009).




\end{thebibliography}
\end{document}